\documentclass[12pt]{article}
\usepackage[dvips]{graphicx}
\usepackage{epsfig}

\title{Disordered cellular automaton traffic flow model:\\
Phase separated state, density waves and self organized criticality}  
\author{K. Fourrate and M. Loulidi\\         
LMPHE, D\'epartement de Physique B.P.1014,\\
Facult\'e des Sciences, Agdal, Rabat, Morocco.}
\begin{document}           
\date{}
\maketitle
\begin{flushleft}
\large\bf{Abstract}
\end{flushleft}
\hspace*{0.5cm}We suggest a disordered traffic flow model that captures many features of traffic flow. It is an extension of the Nagel-Schreckenberg (NaSch) stochastic cellular automata for single line vehicular traffic model. It incorporates random acceleration and deceleration terms that may be greater than one unit. Our model leads under its intrinsic dynamics, for high values of braking probability $p$, to a constant flow at intermediate densities without introducing any spatial inhomogeneities. For a system of fast drivers $p\rightarrow 0$, the model exhibits a density wave behavior that was observed in car following models with optimal velocity. The gap of
the disordered model we present exhibits, for high values of $p$ and random deceleration, at a critical density, a power law distribution which is a hall mark of a self organized criticality phenomena.\\


\newpage
\section{Introduction}
\hspace*{0.5cm}The investigation of traffic flow has attracted the interest of physicists already a long time ago. Different approaches have been proposed[1,2]. One can distinguish macroscopic and microscopic ones. In macroscopic models the traffic is viewed on the one hand as a compressible fluid formed by the vehicles and then analyzed using hydrodynamical fluid theories[3], on the other hand as a gas of interacting particles and then treated using kinetic theories of gases based on the Boltzmann equation[4].\\
\hspace*{0.5cm}In the car-following theories[5 and references therein], which are a typical example of microscopic approaches, individual vehicles are distinguished and the equation of motion, for each one, is the analogue of the Newton's equation. Many car-following models have been proposed depending on a sensitivity parameter and reaction time. But, they lead to unrealistic description of the behavior of free vehicles which have a large distance to the next vehicle ahead and suffer from serious problems in the low -density limit. To overcome these problems an optimal velocity $V_{opt}$ was introduced and a typical form[6] gives satisfactory results compared with empirical data. But, it has been found empirically that $V_{opt}$ depends on the traffic state[7]. For a detailed study of all the above approaches and theories we refer the reader to the review articles Ref 5,8 and references therein.\\
\hspace*{0.5cm}CA approach can be considered as a powerful tool, in statistical physics, to model local and nonlocal interactions[9]. Nowadays,  the simulation of traffic using cellular automata approach[10] stands out for its simplicity. It is a microscopic description of traffic flow governed by simple rules that each individual driver follows.\\
The stochastic traffic CA model introduced by Nagel and Schreckenberg[11] (NaSch) is governed by simple rules. It simulates single-lane one way traffic and is able to reproduce the main features of traffic flow as backward moving shock waves and the so-called fundamental diagram, $J=J(\rho)$. In real traffic the system dynamics is very complex. However, the stochasticity introduced into the model takes into account of some events due to human driving: the maximum speed fluctuations, overreactions at braking and retarded acceleration. The NaSch model has been intensively studied using both analytical and numerical methods[12]. Many extensions of this model have been established in order to understand the rich variety of physical phenomena exhibited by vehicular traffic[13,14]. Some of these phenomena, observed in vehicular traffic under different circumstances, include transitions from one dynamical phase to another, criticality and self-organized criticality, metastability and hysteresis, phase-segregation, etc. We note that no one of these extensions is able to reproduce the main phenomena of real traffic flow all together.\\
\hspace*{0.5cm}Since inhomogeneities have relevant effects on the systems dynamics, many kind of disorder were involved in traffic flow models. It was shown numerically that the NaSch model with a quenched random deceleration probabilities displays queueing of cars with a power law distribution of gaps between the cars at low densities[15]. The jammed phase behavior is similar to that observed in the standard NaSch model. The introduction of defects in NaSch model has a very high impact on the fundamental diagram and the dynamics of the model[16,17]. One distinguishes between two kind of defects, the sitewise and the particlewise disorder. The later case produces phase separated stationary states, which consists of a large jam behind the slowest vehicle and a large gap in front of it at low densities. The single defect site induces a third phase, which is also a phase separating, located between free flow regime and jamming phase where the flow is constant[17]. This intermediate phase was already observed in the asymmetric exclusion model, which corresponds to the NaSch model with $v_{max}=1$, for the sitewise disorder[18].\\
\hspace*{0.5cm}Since the CA approach may give, in a simple way, a good description of real traffic and captures the main phenomena observed by analyzing empirical data[7,19], we suggest an extension of the NaSch model that includes explicitly random incrementation of driver velocities.
However, we think that it should be more realistic to take into account of non uniform acceleration and deceleration of vehicles as the drivers act differently. In the road you find careful drivers, which don't drive fast, as well as careless ones, which drive at their maximum speed. Depending on the distance ahead and the velocity of the vehicle the drivers may accelerate or decelerate more than one unit at each time step. Usually the competition between quenched randomness and dynamic fluctuations induces phase transitions between a disordered-dominated phase and fluctuation-dominated phase with qualitatively distinct behaviors. Thus, we expect that  a model with random acceleration should induce new dynamical states especially for intermediate densities since the fast drivers are stuck by the slowest ones. The model is defined in section 2 and depending either on random acceleration or deceleration we show numerically in section 3 that it presents some interesting phenomena which were observed separately in different varieties of traffic flow models[19-21,23]. In section 4 we study the gap distribution in order to show that the model exhibits a critical self organized behavior. The conclusion of our mean results is given in section 5.

\section{Definition of the model}
\hspace*{0.5cm}The disordered traffic flow model we present is a probabilistic CA where not only space and time are discret, but also the state variable of the vehicles. As in the NaSch model each cell can be empty or occupied by exactly one vehicle $n$ and the state of each one is characterized by its velocity $v_n$ which can take one of the $v_{max}+1$ allowed integer values $v=0,1,2,..,v_{max}$. We denote the position and the velocity of the $n$-th vehicle by $x_n$ and $v_n$  respectively . Then, $g_n=x_{n+1}-x_n-l$, where $l$ is the vehicle length, is the gap between the $n$-th vehicle and vehicle $n+1$ in front of it. At each time step $t \rightarrow t+1$, the $N$ vehicles arrange themselves on a finite lattice of length $L$ following a parallel update according to the following rules:\\
\it{Step 1: Acceleration}.\\
\rm If $v_n < v_{max}$, the velocity of the $n$-th vehicle is increased by $a_n$ sites, i.e.
\begin{equation}
v_n\rightarrow min(v_n + a_n,v_{max})
\end{equation}
where $a_n = [p_ng_n]+1$.\\  
\it{Step 2: Deceleration}.\\
\rm If $(g_n+1)\leq v_n$, the velocity of the $n$-th vehicle is reduced to $g_n$, i.e.
\begin{equation}
v_n\rightarrow min(v_n,g_n)
\end{equation}  
\it {Step 3: Randomization}.\\
\rm If $v_n > 0$, with probability $p$ the velocity of the $n$-th vehicle is decreased randomly by $d_n$ sites, i.e.
\begin{equation}
v_n\rightarrow max(0,v_n-d_n) \hspace*{1.1cm} \mbox {with the probability $p$}
\end{equation}
where $d_n = [q_n min([v_{max}],g_n)]+1$.\\
\it {Step 4: Vehicle mouvement}.\\
\rm Each vehicle moves forward according its new velocity $v_n$ obtained from the steps 1-3, i.e.
\begin{equation}
x_n \rightarrow x_n + v_n
\end{equation}
The symbol [$A$] denotes the integer part of $A$ and the quenched $p_n$ and $q_n$ variables are randomly distributed in the interval $[c,1]$ according to the distribution laws:
\begin{equation}
\varphi(p)=\frac{n+1}{(1-c)^{n+1}}(p-c)^n
\end{equation}
and
\begin{equation}
\psi(q)=\frac{n+1}{(1-c)^{n+1}}(1-q)^n
\end{equation}

As the model we suggest is an extension of the NaSch model, the four steps quoted above are necessary to reproduce the basic features of real traffic[5,13]. In step 1 the driver benefits from all the distance to the vehicle ahead and drives as fast as possible without crossing the maximum speed limit. He might move $[p_n g_n]+1$ sites. In step 2, to avoid collision between vehicles, the driver reduces its velocity and adjusts it according to the distance to the vehicle ahead to be in security from any crash. The different behavioral patterns of the individual drivers as non deterministic acceleration as well as overreaction while slowing down are reflected in step 3. This step is crucial for the spontaneous formation of traffic jams. In real traffic the driver ability to drive more or less fast is a pertinent parameter in the dynamics of vehicles. The distributions given above reflect the fact that the 'careless' drivers drive as fast as possible. They correspond to higher values of $p_n$ and lowest values of $q_n$. On the other hand the 'careful' drivers drive slowly since their corresponding probabilities $p_n$ and $q_n$ are the lowest respectively the highest.

\section{Numerical investigations of the fundamental diagram}
\hspace*{0.5cm}Our numerical investigations are performed on a ring composed of $L$ sites. The density of the vehicles is given by $ \rho = N/L$ where $N$ is the number of vehicles. The flow $<J>$ is expressed as $\rho <v>$ where the mean velocity $<v>$ of the vehicles is defined as the fraction of the sum of movable vehicle velocities among N vehicles. The length of the vehicles, $l$, will be taken as the unit of space coordinate. We use a parallel updating sheme since it takes into account the reaction-time and can lead to a chain of overreactions. In what follow we will present the results obtained essentially from the distribution defined in eq(5,6) for $n=1$. In order to analyze the effects of different random variables we will discuss 3 cases:

\subsection{\it deceleration effects}
\hspace*{0.5cm}We suppose that our vehicle drivers are all "careful" and they drive as slow as possible. They accelerate just by one unit at each time step while in the breaking step 3 they may reduce their velocity as maximum as possible. The fundamental diagram in this case, which corresponds to $p_n=0$ for all vehicles, depends on the randomization parameter values $p$. However weak the value of $p$ is, the flow, which usually presents a sharp maximum, exhibits a smooth variation at its maximum value (Fig. 1a). This is due to the fact that the mean velocity of vehicles, $v=v(\rho)$, is a decreasing function of $\rho$, even at low densities(Fig. 1b).\\

\begin{figure}
\begin{center}
\includegraphics[width=8cm,height=6cm,angle=0]{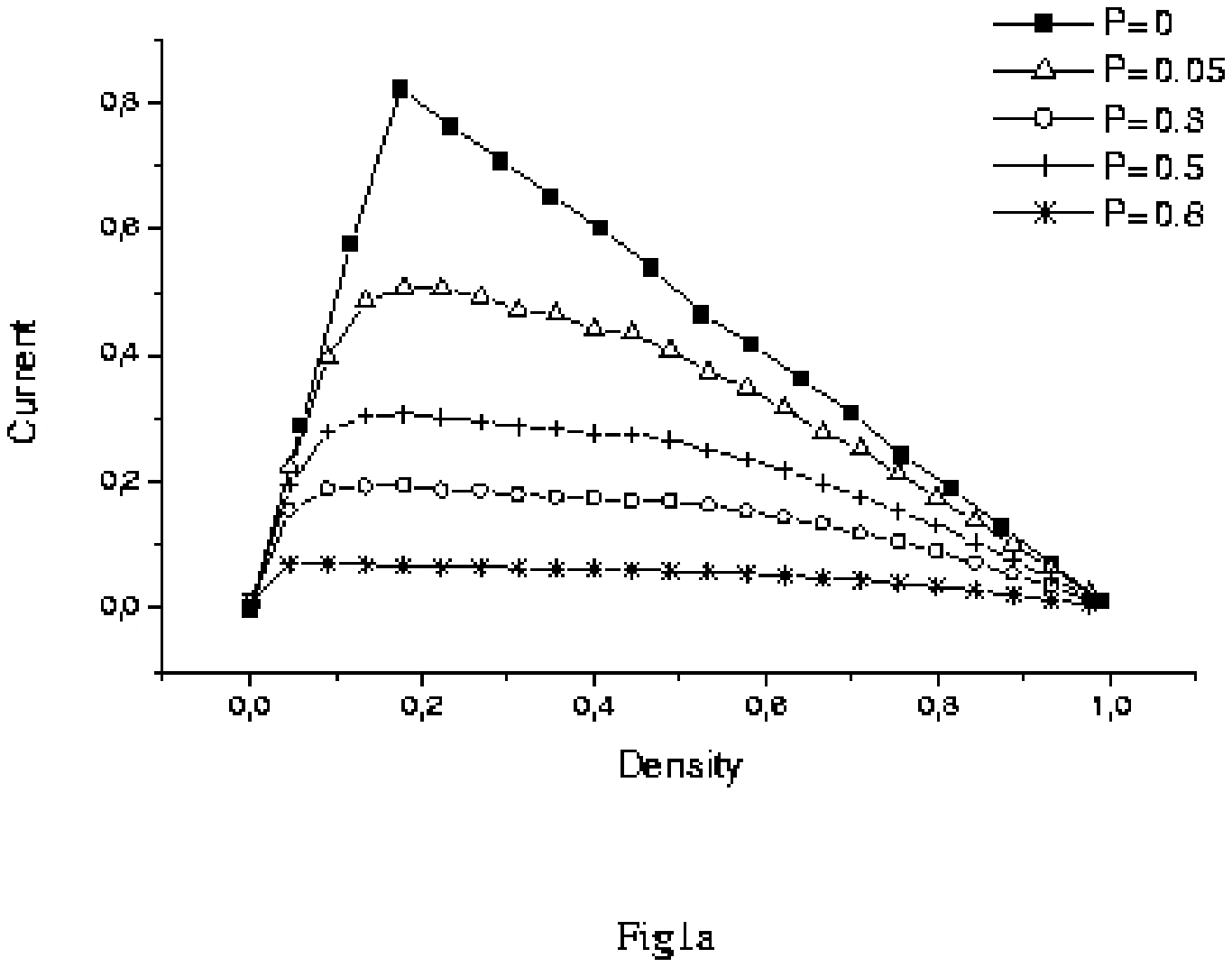}
\includegraphics[width=8cm,height=6cm,angle=0]{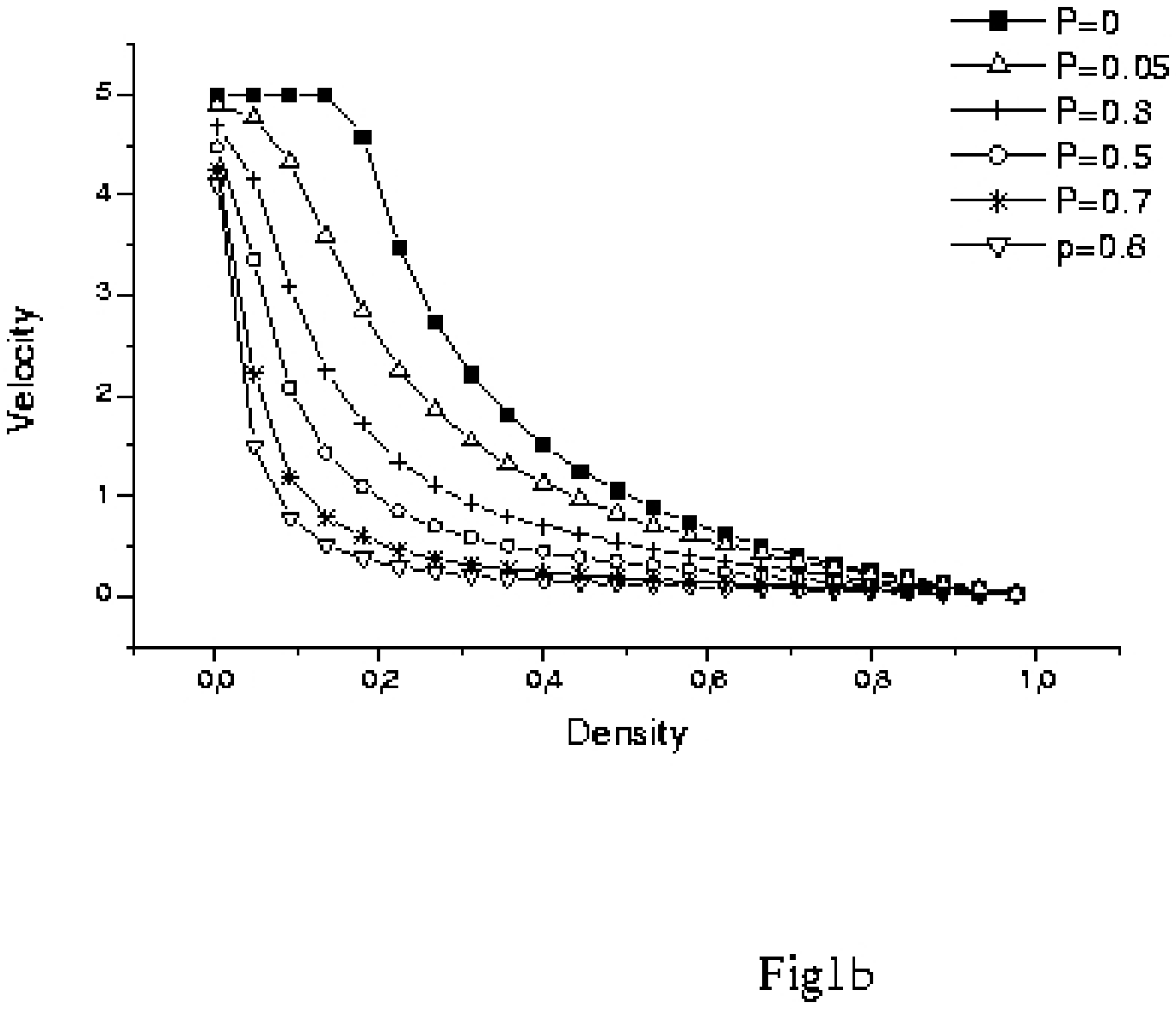}
\caption{(a)The fundamental diagram of the model and (b) $\rho$-dependence of the velocity for $v_{max}=5$ in the case of random decelerations and
uniform acceleration.}
\end{center}
\end{figure}

\hspace*{0.5cm}The existence of very slow vehicles in the system induces a jamming regions even at very low densities. The fast vehicles pile up behind the slowest ones. As a matter of fact, the space-time diagram presents, for intermediate values of $p$, a macroscopic high density region confined between relatively free flow ones(Fig. 2a). This results from the fact that the slowest vehicles, which are randomly distributed in the system, act like a blockage. Thus, the flow gets a slower value than in the Nasch model. By increasing the global density within the segregated phase the bulk densities of these regions remain constant, only their length changes. Consequently, the average flow is constant in the segregated phase, because the average density in the vicinity of the high density region does not depend on the global density. This intermediate phase, which is located for $\rho_{low}<\rho<\rho_{high}$ may be illustrated looking at the density profile which is plotted in fig 2b. For any particular choice of random $q_n$ variables according to the distribution law $g(q)$(eq. 6) one observes a separation into macroscopic high and low density regions. As the disorder average wasn't taken into account the density fluctuations observed in both regions
result from the random decelerations of vehicles. They may also result from the interaction between vehicles that cause a formation of local jams (gaps) in the 'free' flow(congested)region.\\

\begin{figure}
\begin{center}
\includegraphics[width=8cm,height=6cm,angle=0]{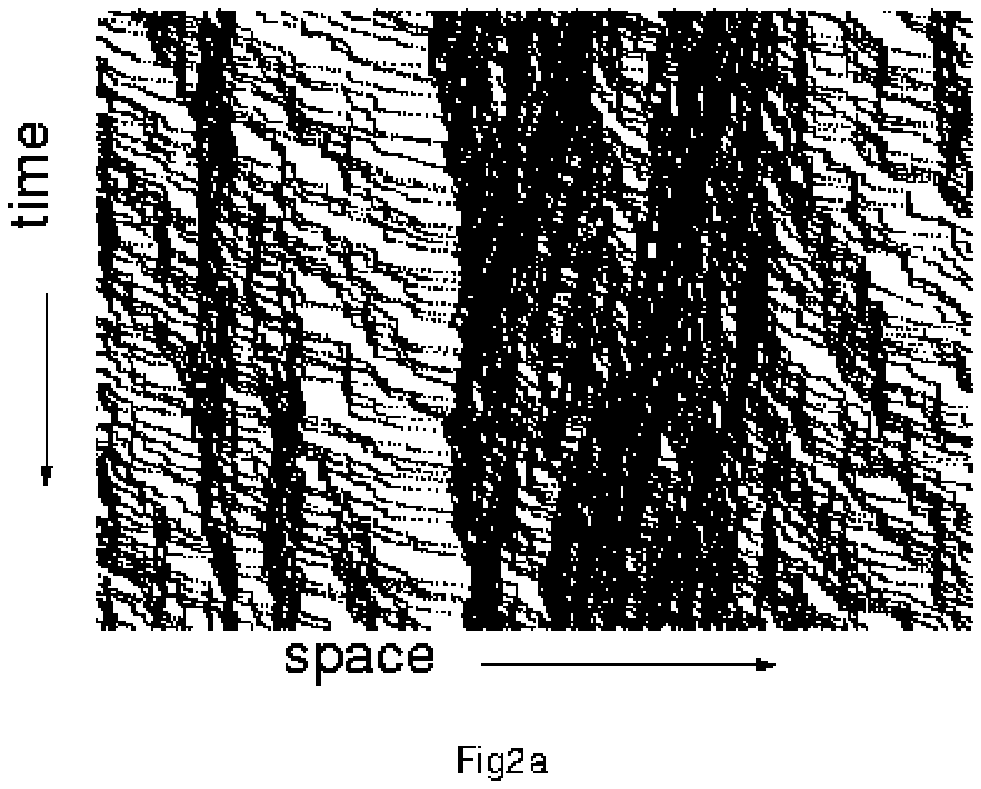}
\includegraphics[width=8cm,height=6cm,angle=0]{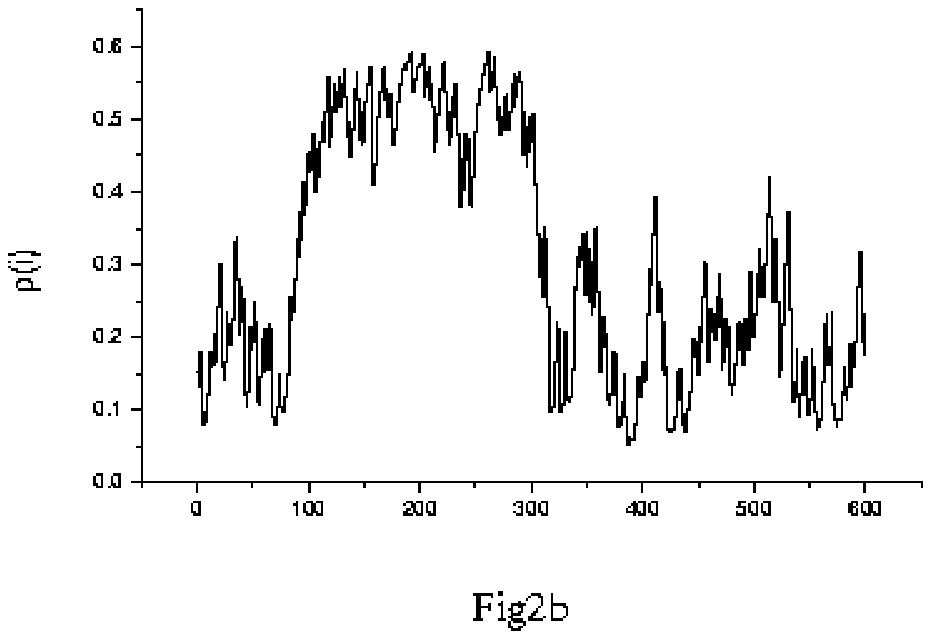}
\caption{(a)The space-time diagram of the model and (b) the corresponding site-dependence of the density, in the case of random decelerations and uniform acceleration, for $\rho=0.3$ and randomization parameter $p=0.8$.}
\end{center}
\end{figure}

 Thus, the average on the disordered $q_n$ variables leads to a quasi-constant flow in the segregated phase(Fig 3). As a result, one can distinguish, in the fundamental diagram, three different phases depending on the global density namely low density phase, high density phase and intermediate phase for $\rho_{low}<\rho<\rho_{high}$. By increasing $\rho$ in this intermediate phase, the flow presents a slow linear decrease. Its slope which is a decreasing function of $p$ vanishes for some randomization parameter values.\\
This phase separation located in our model was found in systems with ramps and systems with a stationary defect[8,19]. For intermediate densities $\rho_{low}<\rho<\rho_{high}$ the flow is constant for these models. In this regime $J(\rho)$ is limited either by the capacity of the ramp or the defect site. Even though the perturbations in these models are different, they all exhibit a plateau formation in the fundamental diagram as well as phase separation in the system independently of the nature of blockage. In the case of ramps it is the local increase of the density that decreases the flow locally. In the model with a stationary defect the increased slowdown parameter leads to a local decrease of the flow. In our model the "careless" drivers decelerate more than one unit and they act as a blockage in the system. Thus, the increased slowdown vehicles leads to a local decrease of the flow.\\

\begin{figure}
\begin{center}
\includegraphics[width=8cm,height=6cm,angle=0]{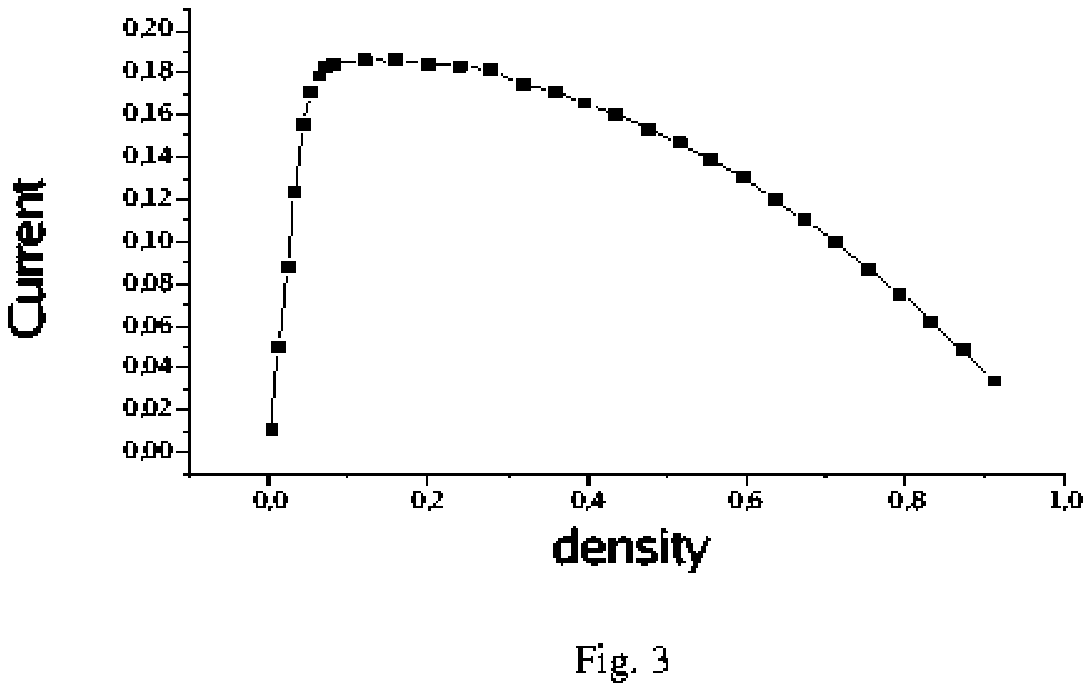}
\caption{The fundamental diagram for $p=0.6$ showing a plateau at intermediate 
densities.}
\end{center}
\end{figure}

\hspace*{0.5cm}Other variants of the model defined above may be obtained by modifying step2 or both step2 and step3 in the dynamics rules (eq. 2-3). If we assume that in the second step the vehicles may decelerate by reducing their velocities to be at a distance to the vehicle ahead more large than one unit such that $v_n\rightarrow min(v_n,g_n+1-d_n)$, the model presents a plateau formation in the fundamental diagram even for very low values of the randomization parameter $p$. The same result is obtained by changing both step2 and step3 of the model rules:
\begin{equation}
\begin{array}{cc}
\mbox{step2:} \hspace*{0.5cm}v_n\rightarrow min(v_n,g_n+1-d_n)\\
\mbox{step3:} \hspace*{0.5cm}v_n\rightarrow max(0,v_n-1) & \mbox{with a probability $p$}
\end{array}
\end{equation}
Since the deceleration effects in these model variants  are emphasized, a plateau formation in the fundamental diagram occurs even at lower values of the randomization parameter $p$.

\subsection {\it Acceleleration effects}
\hspace*{0.5cm}Even though that the acceleration of vehicles is over estimated we think that it is worthwhile to study the effects of randomness of a such parameter. Those effects on the fundamental diagram and on the dynamics of the model are studied by setting $q_n=0$ and choosing randomly the variables $p_n$ for all vehicles. In this situation all the drivers are 'careless' and drive as fast as possible. By performing numerical investigations we show that, for low values of $p$, the fundamental diagram(Fig. 4a) has quite similar form to that one obtained in the NaSch model. Thus, the distinction between these models should be formulated at microscopic level.\\
As usual, the fundamental diagram depends on the randomization parameter $p$. For low values of $p$ it presents two different regimes with respect to the density. At low density the system is in a free flow state and all the vehicles move, up to a certain density $\rho_1$, at their maximum speed (Fig 4b). By increasing $p$ this free flow region decreases considerably , i.e. $\rho_1$ decreases, because the vehicles may frequently break and then reduce their speed more than one unit. This leads to the formation of spontaneous jams that are scattered in the space time diagram(Fig. 5a). By increasing the density more jams arise and are grouped together leading to a large strip of jammed region separated by free flow regions(Fig. 5b).\\
Depending on the randomization parameter $p$ the microscopic state of the system exhibits different structures. Indeed, for low values of $p$ the system may selforganize into a density wave state(Fig. 6). For a such values of $p$ most of the vehicles accelerate randomly and benefit from the distance of the vehicle ahead. In addition, the velocity of all vehicles in jammed region is limited by the slowest one. Then, if most of the vehicles located at the jammed region drive fast(slow)we have $J_{in}<J_{out}$ ($J_{in}>J_{out}$), where $J_{in}$($J_{out}$) denotes the number of entring(leaving) vehicles in the jammed region per unit of time. As the incoming vehicles pile up behind the jam we have $\rho_{out}<\rho_{in}$ where $\rho_{out}$($\rho_{in}$) is the density of vehicles that leave(drive in) the jam. Consequently, the domain-wall between two stationary regions, free flow and congested region, propagates either in the opposite or the same direction of moving vehicles with the velocity $v_{shock}$, which depends on the values of $J_{in}$ and $J_{out}$. From mass conservation we obtain

\begin{equation}
v_{shock}=\frac {j_{out}-J_{in}}{\rho_{out}-\rho_{in}}
\end{equation}

\begin{figure}
\begin{center}
\includegraphics[width=8cm,height=6cm,angle=0]{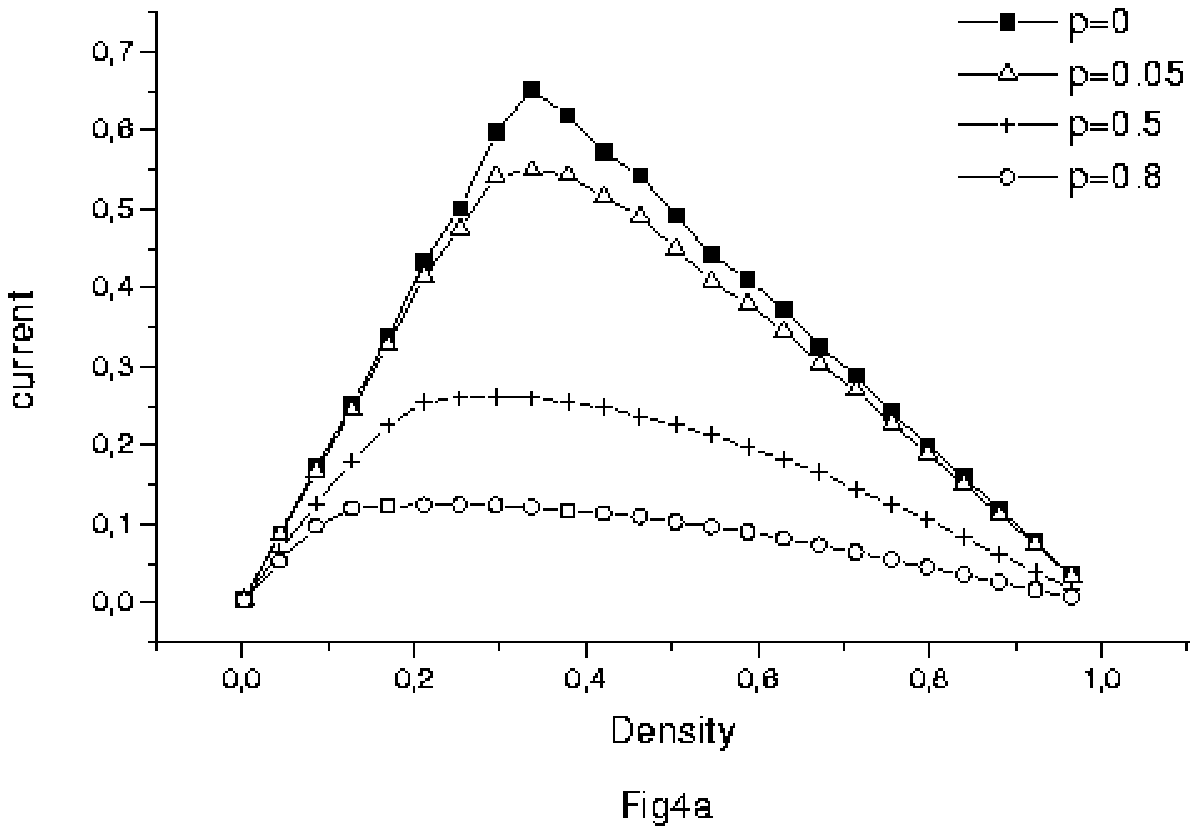}
\includegraphics[width=8cm,height=6cm,angle=0]{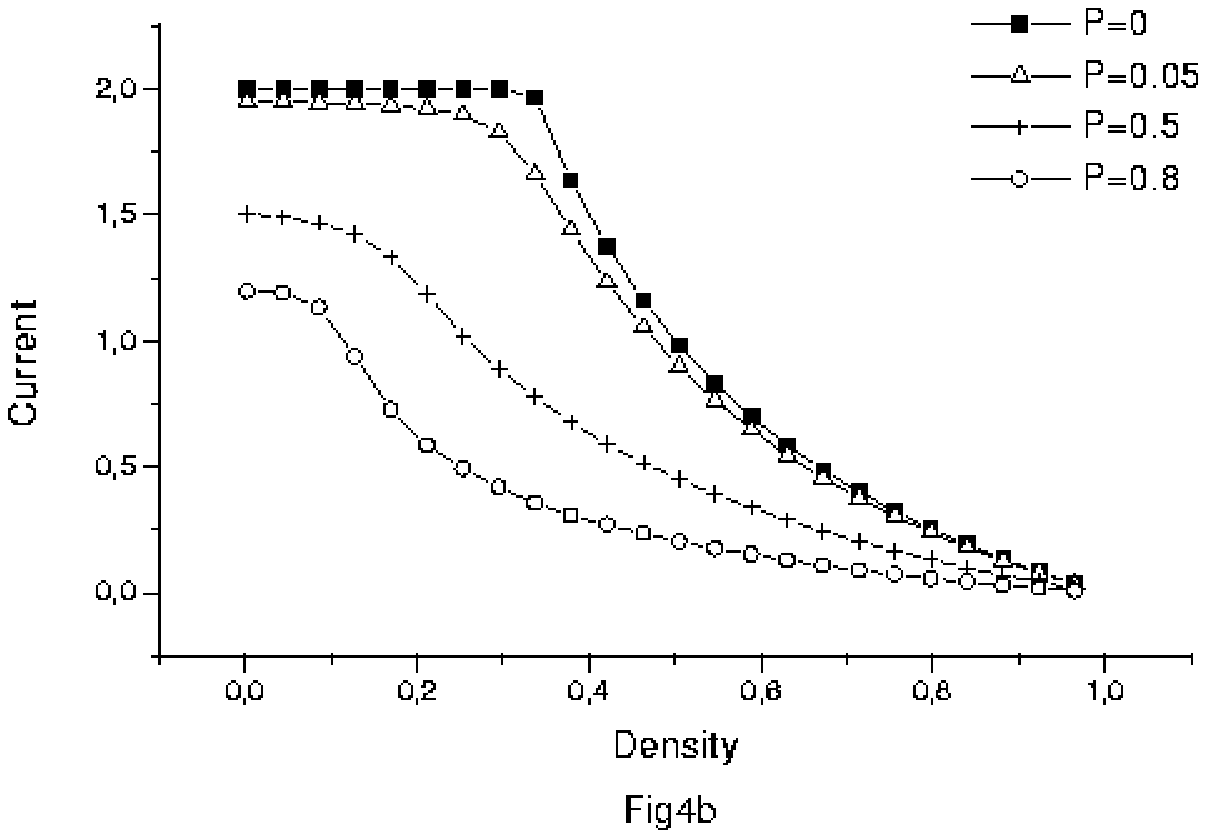}
\caption{(a)The fundamental diagram of the model and (b)$\rho$-dependence of 
the velocity for $v_{max}=2$ in the case of random accelerations and 
uniform deceleration.}
\end{center}
\end{figure}

As a result, the jammed regions wave from side to side. A such behavior was observed in car following models with an optimal velocity that takes into account the characteristics of different vehicles[20]. In those models the different maximum velocities among vehicles are included in the optimal velocity function while in our CA model all vehicles keep the same $v_{max}$ and the characteristics of different drivers are induced in the acceleration terms. We note that the model presents regular oscillating behavior (Fig.  6a) for vanishing value of the randomization parameter, i.e. $p=0$, while for finite values of $p$ the breaking rule induces some irregularities. Detailed numerical investigations of the phase transition among the freely moving phase, the density wave phase and the homogeneous congested phase show that there is a critical value of the randomization parameter, $p_c$ above which the density wave does not appear[21]. If we denote by $g_f$ ($g_i$) respectively $v_f$ ($v_i$) the average headway and velocity within (out of) the jam we may introduce an order parameter $s=g_f - g_j$ or $s=v_f -v_j$ and study the dynamics of the jammed region and the critical behavior of this inhomogeneous phase.

\begin{figure}
\begin{center}
\includegraphics[width=8cm,height=6cm,angle=0]{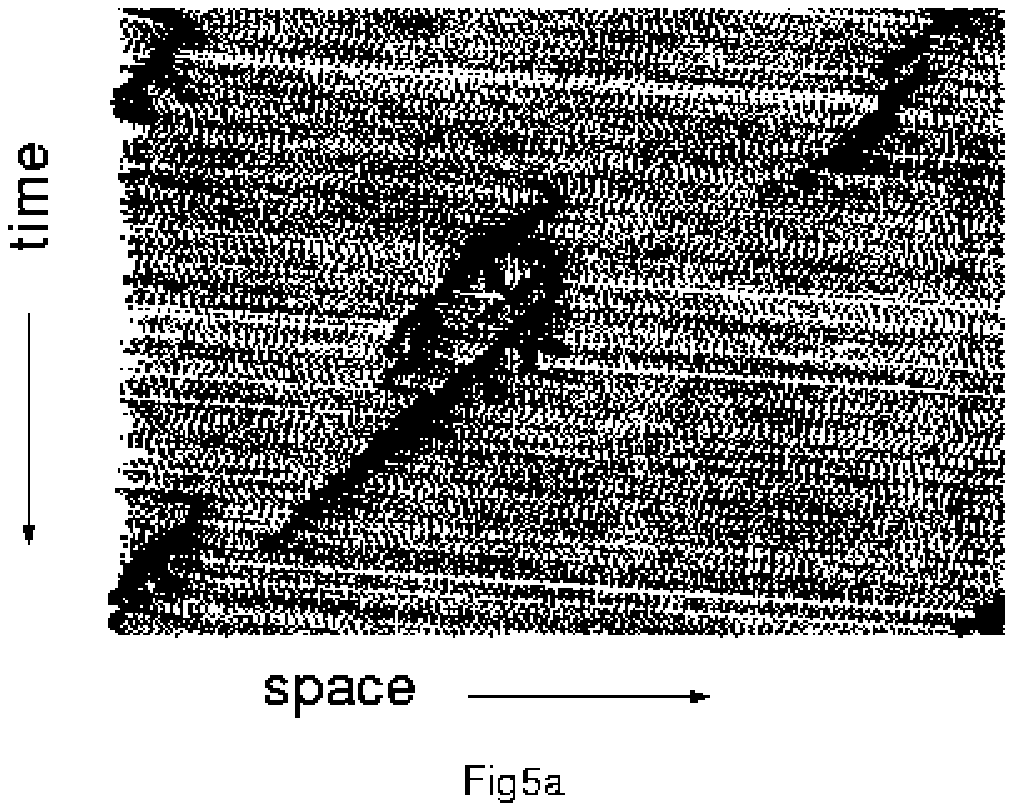}
\includegraphics[width=8cm,height=6cm,angle=0]{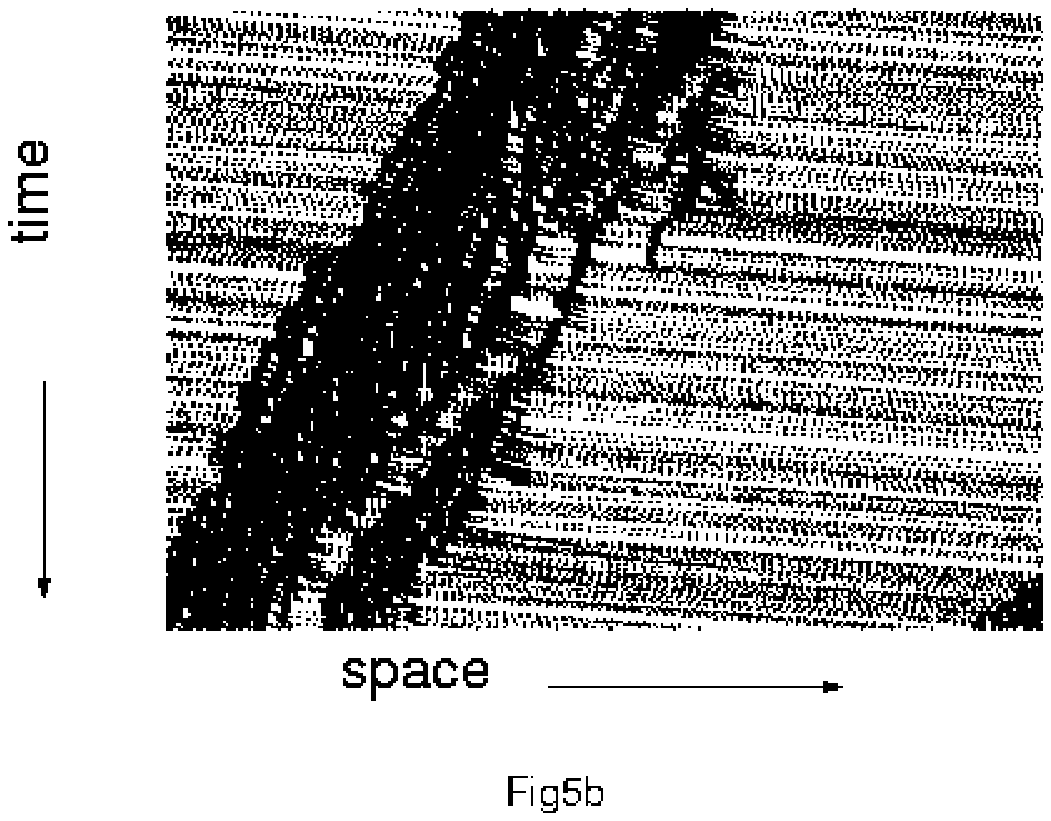}
\caption{The space-time diagram of the model,(a)$\rho=0.1$ and (b)$\rho=0.2$, in the case of random acceleration and uniform decelerations for $v_{max}=5$ and $p=0.8$.}
\end{center}
\end{figure}

\subsection{\it Acceleration and deceleration effects}
\hspace*{0.5cm}In real traffic we found both "careful" and "careless" drivers. Then, both variables $p_n$ and $q_n$ should be chosen randomly. For very low values of randomization parameter $p$, most of vehicles drive fast. Consequently, at intermediate densities the steady state of the system reaches a density wave state with local 'defects' that are caused by spontaneous jams. This microscopic structure results from the interaction between the fast vehicles and the slowest ones.\\
As in the case of slow drivers, i.e $p_n=0$, a region of constant flow occurs for high values of $p$($p>\frac{1}{2}$). It becomes larger than in the previous case. Indeed, the faster drivers are blocked by the slowest ones and they pile up behind the jammed region. Consequently, the system exhibits, at intermediate densities, a high density band confined between relatively free flow regions.

\begin{figure}
\begin{center}
\includegraphics[width=8cm,height=6cm,angle=0]{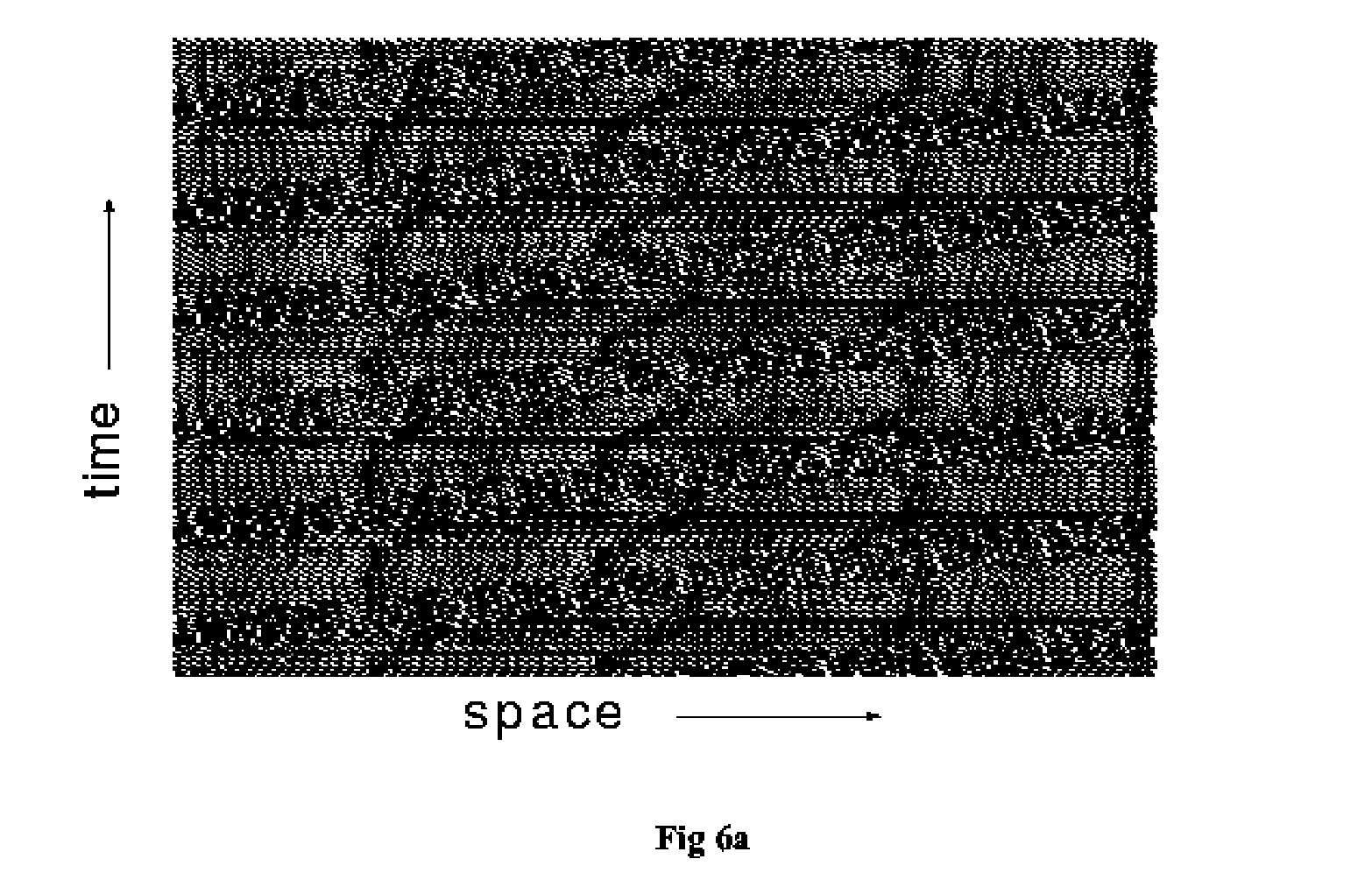}
\includegraphics[width=8cm,height=6cm,angle=0]{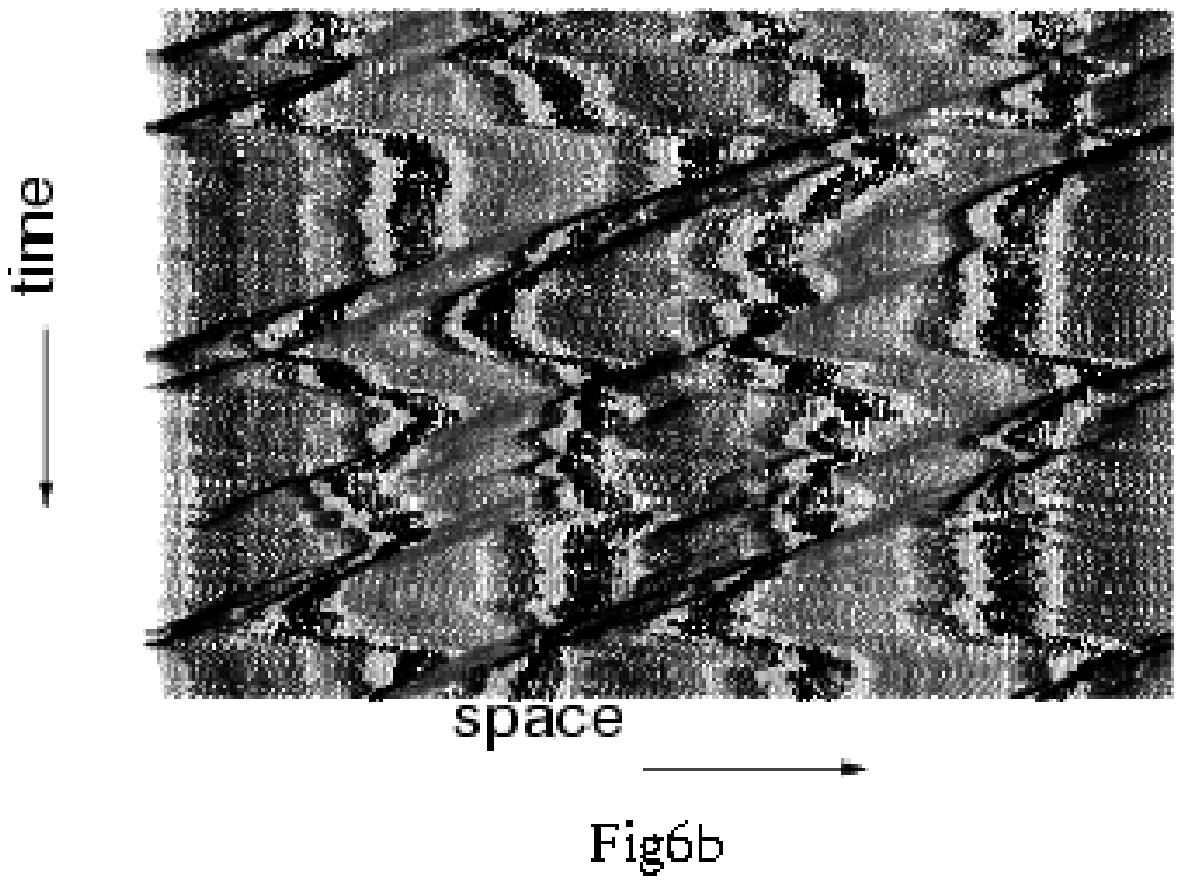}
\includegraphics[width=8cm,height=6cm,angle=0]{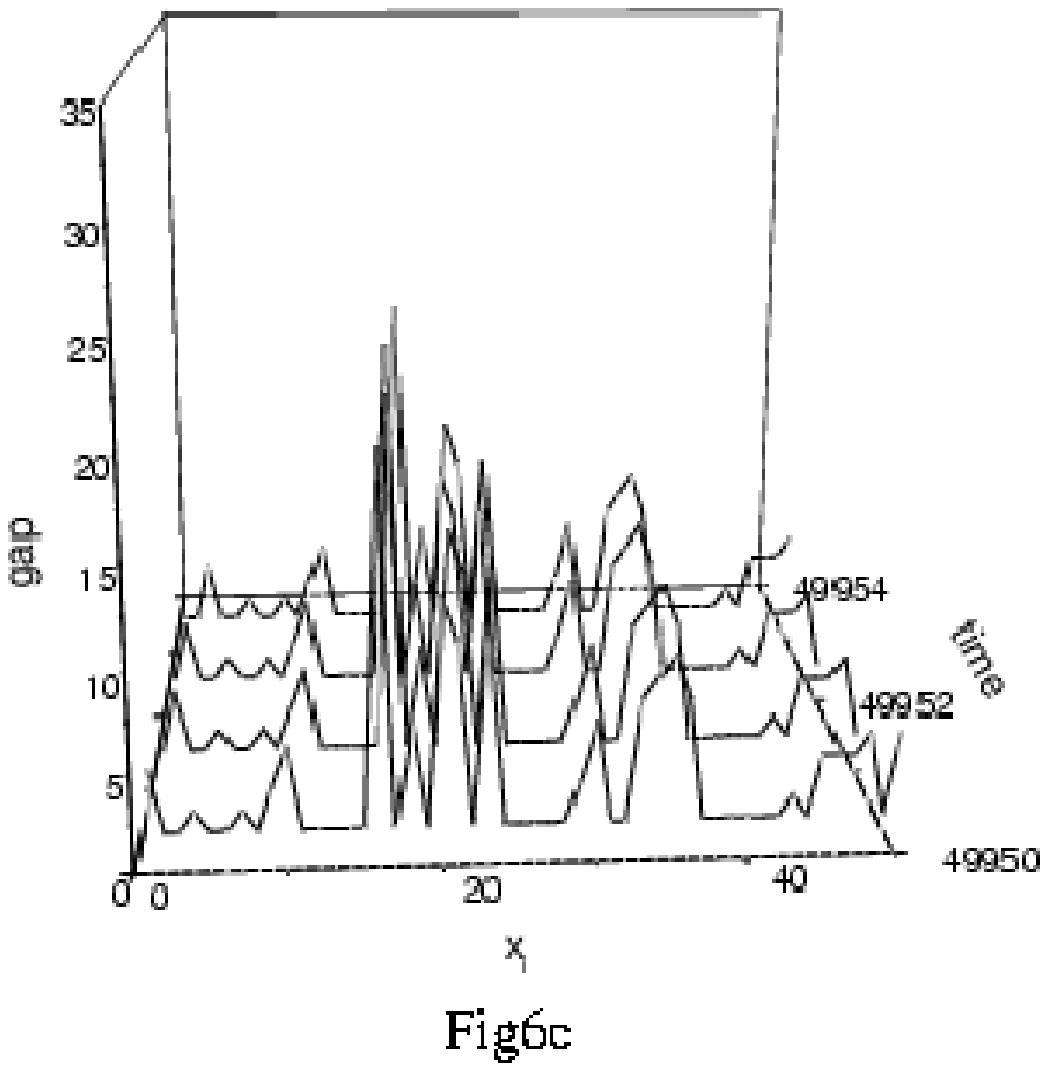}
\caption{The space-time diagram of the model in the case of random decelerations and uniform acceleration for $v_{max}=5$, (a)$\rho=0.15$ and $p=0$, (b) $\rho=0.2$ and $p=0.05$ and (c)the spatio-temporal evolution of the gap, for $\rho=0.2$ and $p=0.05$. A density wave behavior is detected.}
\end{center}
\end{figure}

\section{The Gap distribution of the model}
\hspace*{0.5cm}Looking at the gap distribution over a wide range of densities in the congested flow region we see that it depends on the maximal velocity $v_{max}$ and the randomization parameter $p$. In the case of high values of $v_{max}$ it exhibits, for low values of $p$, two maxima (Fig. 7). The position of the maxima do not change considerably over a wide range of densities. The first maximum is assumed for the value zero. This means that the steady state of the system in the congested phase is of a densely packed queue and the two maxima of the gap distribution may be associated simply to a free flow phase with random distributed local inhomogeneities that are densely packed jams. We note that in a continuum limit of the NaSch model[22] the first maximum was observed at a non vanishing value and such a behavior was associated to a phase separation into congested and free flow regions. By decreasing the density, the first maximum vanishes and the gap distribution exhibits only one maximum that corresponds to the free flow regime.\\

\begin{figure}
\begin{center}
\includegraphics[width=8cm,height=6cm,angle=0]{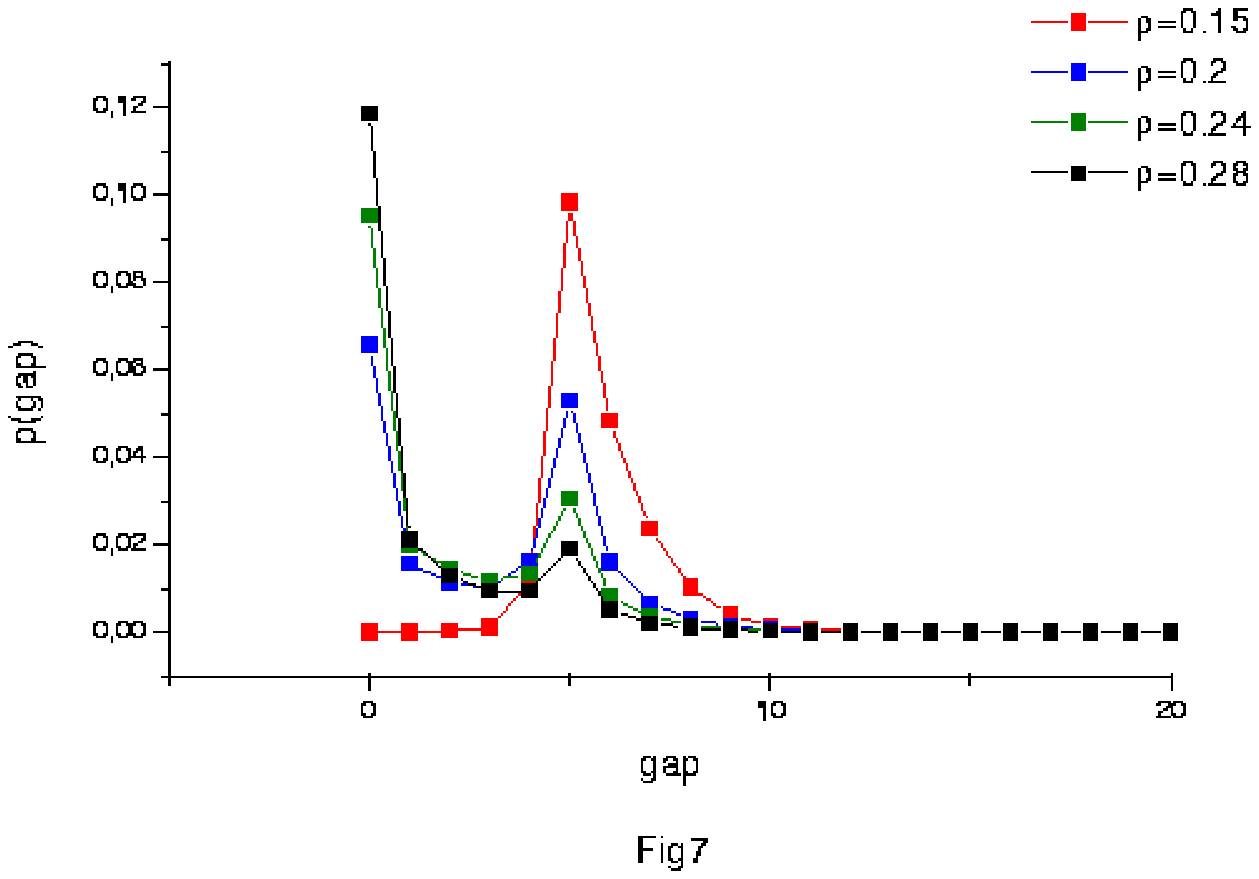}
\caption{The gap distribution of the model, in the case of random decelerations and accelerations, for $v_{max}=2$, $p=0.05$ and different values of $\rho$.}
\end{center}
\end{figure}

For high values of $p$ located in a narrow interval, i.e. $p_{c_1}<p<p_{c_2}$, the model displays, in the case of careful drivers, a power law gap distribution in the vicinity of some critical density(Fig. 8). This behavior is a feature of the self organized criticality (SOC)[23]. 
The distribution doesn't change considerably with the system size and finite size data collapse of the form

\begin{equation}
P(g)=g^{-\tau}f(g/L^{\nu})
\end{equation}

Using a finite size analysis it's easy to see that $P(g)=L^{-\beta}f(g/L^{\nu})$ with $\beta/\nu=\tau$. The critical exponents $\tau$ and $\nu$ vary continuously with the randomization parameter $p$ and the maximum velocity $v_{max}$. The SOC behavior results from the fact that the fast vehicles built behind the slowest ones forming platoons separated by large gaps. The power law decay of the gap distribution was illustrated analytically for the asymmetric exclusion model with random rates using some statistical estimates[24]. The numerical estimation of the critical densities and the critical exponent $\tau$ of the gap distribution of our model are represented in table 1 for $v_{max}=2$ and $v_{max}=5$. We note that the SOC behavior was also observed in a stochastic traffic model with random deceleration probabilities[15]. It was pointed out that the NaSch model with random randomization parameter probabilities self organizes into a stable queueing phase at low densities and has a power law gap distribution. This model belongs to the same universality class of the BFL model[25] while our model belongs to another universality class as the critical exponents $\tau$ are rather different.\\

\begin{center}
\begin{tabular}{|l|r|c|r|}\hline
\multicolumn{1}{|c|}{$v_{max}$} & \multicolumn{1}{|c|}{$p$} &
\multicolumn{1}{c|}{$\rho_c$} & \multicolumn{1}{|c|}{$\tau$}\\ \hline
2	& 0.7	& 0.19	& 0.755 \\ \cline{2-4}
	& 0.8	& 0.18	& 0.812 \\ \hline
5	& 0.75	& 0.1	& 0.915 \\ \cline{2-4}
	& 0.8	& 0.11	& 0.972 \\ \hline
\end{tabular}
\end{center}

\begin{table}[tbh]
\vspace{0.2cm}
\caption{The estimations of the critical exponents $\tau$ of gap distributions at critical densities $\rho_c$ for different $v_{max}$ and different values of the randomization parameter $p$.}
\end{table}

\begin{figure}
\begin{center}
\includegraphics[width=8cm,height=6cm,angle=0]{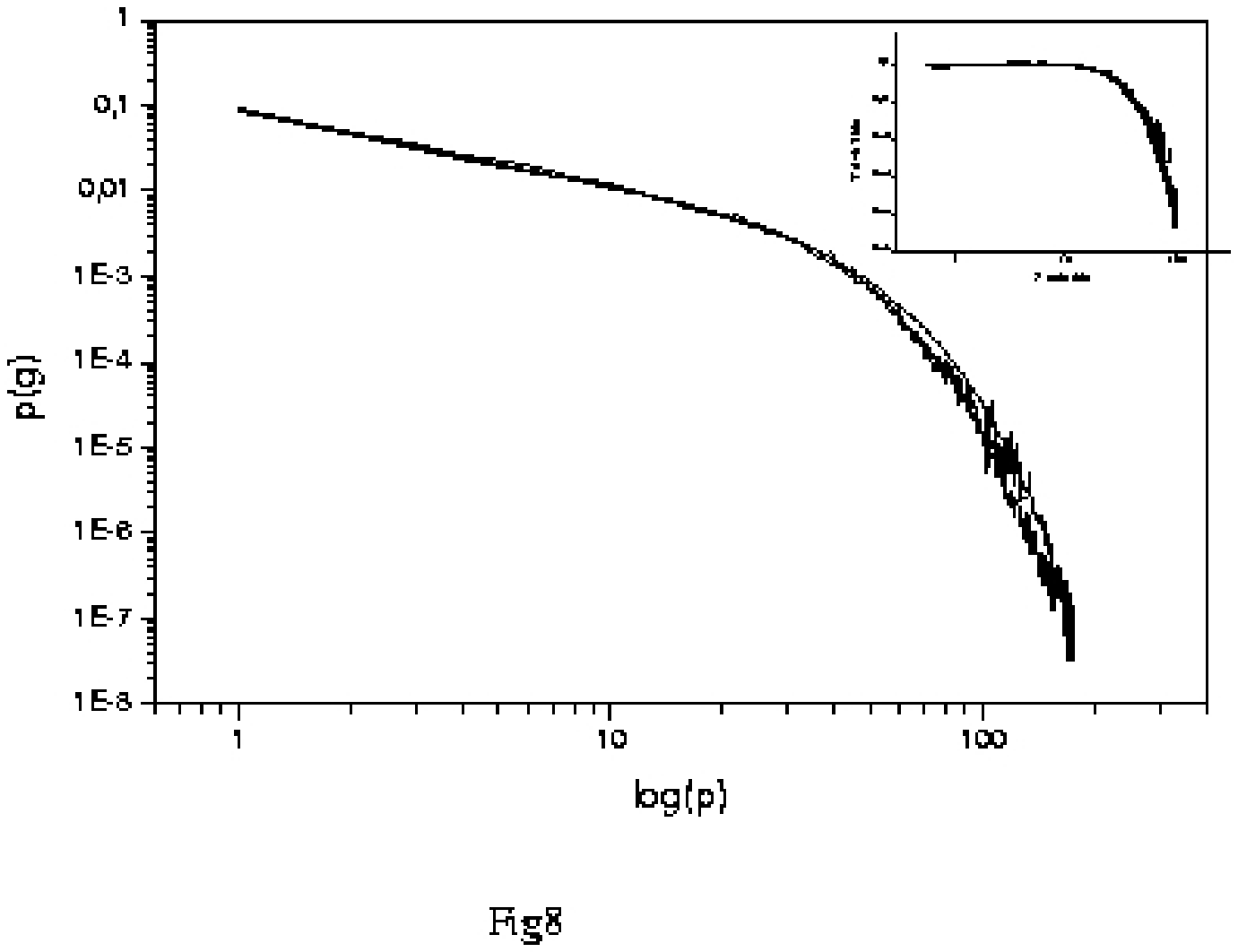}
\caption{The log-log plot of the gap distribution for $v_{max}=2$, $\rho_c=0.1$ and $p= 0.75$, in the case of random accelerations and uniform 
deceleration, and for different system sizes $L$ ranging from $L=860$ to 
$L=3440$. The finite size scaling analysis gives a good fit of our data 
with the exponents $\tau=0.903$ and $\nu=0.083$. The inset shows the finite 
size fit with $\beta=0.075$.}
\end{center}
\end{figure}

In the case of 'careful' drivers, all vehicles accelerate just by one unit at each time step but they may decelerate by reducing their velocity as maximum as possible. The probability that one driver decelerates as maximum as possible is realized for $1-\delta<q_{n}<1$, where $\delta << 1$ and is calculated from the distribution given in eq (6): $p_m \sim \delta^2/2$. At low densities the system evolves according to its own dynamics and is however interrupted by relatively small perturbations at vanishingly small rates. Consequently, the system gets enough time to organize itself and relax back to its corresponding steady state before it is perturbed again. This effectively separates completely the time scales for perturbing the system and its response. This infinite time scale separation is required by SOC behavior. The system is then self similar and the scale invariance is due to the fact that the jams are fractal in the sens that there are smaller sub-jams inside large jams.

\section{Conclusion}
\hspace*{0.5cm}The dynamics of the disordered traffic flow, which is based on the NaSch model, presented in this paper bears considerable structures and many of interesting features of traffic flow. Random acceleration and deceleration terms were introduced in the CA rules. Within numerical investigations we have shown that the model presents under its own dynamics many interesting phenomena that were observed separately in different varieties of traffic flow models namely the segregated phase, the density wave state and the SOC behavior. Depending on the randomization parameter $p$ a rich variety of fundamental diagrams is obtained. For intermediate values of $p$ and low densities the fundamental diagram may trigger transitions from free to congested flow under its own dynamics without introducing any external perturbation like defects or on and off-ramps[7,13]. We may assume that at intermediate densities the slow vehicles act like a defect site leading to a blockage. As a matter of fact an intermediate phase, where the current is quasi-constant, is inserted between free flow and jamming phases. The analysis of the microscopic structure of the space time diagram for low values of $p$ shows that the system may organize itself into a wave density phase where the jams (gaps) oscillate. Such behavior was pointed out by Nagatani in the optimal velocity model[20]. The gap distribution of our model in the case of slow drivers exhibits, for some values of randomization parameter $p$ and at low densities, a power law behavior in the vinicity of a critical density signaling the existence of SOC phenomenon. The critical exponents are rather different from those obtained in a disordered traffic flow model where the quenched disorder was incorporated in the random deceleration probabilities[18]. The model we suggest doesn't belong to the same universality class of the BFL model as the former model does. Finally, we note that the disorder introduced in our model is the essential ingredient to observe all the traffic features mentioned above since they don't occur in the case of uniform acceleration and deceleration[26],i.e $p_n=cst$ and $q_n=cst$.\\
\hspace*{1cm}The model we present in this paper presents many interesting features of traffic flow and may be developed in order to capture other pertinent phenomena observed in traffic flow as the hysteresis and the synchronized state.\\

{\bf Aknowlendgements}\\
\hspace*{1cm}We thank A. Schadschneider for his critical reading the manuscript and for his relevant comments and suggestions. The authors are grateful to the high education ministry MESFCRS for the financial support in the framework of the program PROTARSIII, grant no:D12/22.
   
\newpage    
\section*{References}
\begin{enumerate}
\item [{[1]}]Traffic and Granular Flow,edited by D. E. Wolf, M. Schreckenberg and A. Bachem (World Scientific, Singapore, 1996).
\item [{[2]}] {\it Traffic and Granular Flow '97'}, edited by M. Schreckenberg and D. E. Wolf, (Springer, Singapore, 1998).
\item [{[3]}]K. Nagel, Phys. Rev. E {\bf 53}, 4655, (1996) and references therein.
\item [{[4]}]I. Prigogine and R. Herman, {\it Kinetic theory of vehicular traffic}, (American Elsevier, New York, 1971).
\item [{[5]}]D. Chowdhury, L. Santen and A. Schadschneider, Phys. Rep. {\bf 329}, 199 (2000).
\item [{[6]}]T. Nagatani, Physica  {\bf A253}, 353(1998).
\item [{[7]}]L. Neubert, L. Santen, A. Schadschneider and M. Schreckenberg, Phys. Rev. E{\bf 60}, 6480 (1999).
\item [{[8]}]D. Helbing, Rev. Mod. Phys.{\bf 73} (2001).
\item [{[9]}]A. Benyoussef, N. El Hafidallah, A. El Kenz, H. Ez-Zahraouy and M. Loulidi, Physica {\bf A322}, 506-520(2003); N. Boccara, {\it Automata networks models of interacting populations} in Cellular Automata, Dynamical Systems, Neural Networks,eds. E.Goles and S.Martinez (Kluwer Dordrecht, 1994); P. Bak, C. Tang and K. Wiesenfeld, Phys. Rev. Lett. {\bf 59}, 381 (1987); Phys. Rev. A {\bf 38}, 364 (1988).
\item [{[10]}]S. Wolfram, {\it Theory and applications of Cellular automata}, (World Scientific, 1986); {\it Cellular Automata and Complexity}
(Addison-Wesly, 1994).
\item [{[11]}]K. Nagel and M. Schreckenberg, J. Phys. (France) {\bf I2}, 2221 (1992).
\item [{[12]}]A. Schadschneider, Eur. Phys. J. B{\bf 10}, 573 (1999).
\item [{[13]}]A. Schadschneider, Physica A {\bf 285}, 101 (2000); R. Barlovic, L. Santen, A. Schadschneider and M. Schreckenberg, Eur. Phys. J. B {\bf 5}, 793 (1998); A. Schadschneider and M. Schreckenberg, J. Phys. A{\bf 30}, L69 (1997); J. Phys. A{\bf 31}, L225 (1998); M. Schreckenberg, A. Schadschneider, K. Nagel and N. Ito, Phys. Rev. E{\bf 51}, 2939(1995).
\item [{[14]}]W. Knospe, L. Santen, A. Schadschneider and M. Schreckenberg, Physica {\bf A265}, 614 (1998); A. Pottmeir, R. Barlovic, W. Knospe, A. Schadschneider and M. Schrekenberg, Physica A{\bf 308}, 471 (2002).
\item [{[15]}]D. V. Ktitarev, D. Chowdhury and D. E. Wolf, J. Phys. A{\bf 30}, L221 (1997).
\item [{[16]}]W. Knospe, L. Santen, A. Schadschneider and M. Schreckenberg, Physica {\bf A265}, 614 (1998).
\item [{[17]}]A. Pottmeir, R. Barlovic, W. Knospe, A. Schadschneider and M. Schrekenberg, Physica {\bf A308}, 471 (2002).
\item [{[18]}]M. Bengrine, A. Benyoussef, H. Ez-Zahraouy and F. M'hirech, Phys. Lett. {\bf A253}, 135 (1999); G. Tripathy and M. Barma, Phys. Rev. E{\bf 58}, 1911 (1998); H. Emmerich and E. Rank, Physica {\bf A216}, 435 (1995); S. A. Janowski and J. L. Lebowitz, J. Stat. Phys. {\bf 77}, 35 (1994).
\item [{[19]}]B. S. Kerner and H. Rehborn, Phys. Rev. E{\bf 53}, R4275 (1996).
\item [{[20]}]T. Nagatani, Physica A {\bf 284}, 405 (2000), Phys. Rev. E {\bf 61},3564 (1999).
\item [{[21]}]K. Fourrate and M. Loulidi, work in progress
\item [{[22]}]S. Krauss, P. Wagner and C. Gawron, Phys. Rev. E{\bf 54}, 3707 (1996).
\item [{[23]}]P. Bak, C. Tang, and K. Wiesenfeld, Phys. Rev. A {\bf 38}, 364 (1988); C. Tang and P. Bak, Phys. Rev. Lett. {\bf 60}, 2347 (1988);
J. Stat. Phys. {\bf 51}, 797 (1988); P. Bak and K. Chen, Physica D {\bf 38}, 5 (1989); K. Wiesenfeld, C. Tang, and P. Bak, J. Stat. Phys. {\bf 54}, 1441 (1989); P. Bak, {\it How nature works: the science of self-organized criticality}. - (Springer Verlag,1996); D. Dhar, e-print cond-mat/9909009.
\item [{[24]}]J. Krug and P. A. Ferrari, J. Phys.A{\bf 29}, L465 (1996); J. Krug, Braz. J. Phys. {\bf 30},97(2000)(cond-mat/9912411);
M. R. Evans, Europhys. Lett. {\bf 36},13 (1996).
\item [{[25]}]I. Benjamini, P. A. Ferrari and C. Landim, Stoch. Proc. Appl. {\bf 61}, 181 (1996).
\item [{[26]}]K. Fourrate and M. Loulidi, unpublished.
\end{enumerate}

\end{document}